\documentclass[journal=jacsat,manuscript=article]{achemso}

\usepackage{textcomp,gensymb,amsmath,xcolor,physics}
\usepackage{pdfpages}
\usepackage[normalem]{ulem}

\newcommand{\txr}{\textcolor{black}}
\newcommand{\txc}{\textcolor{black}}
\newcommand{\txb}{\textcolor{black}}

\author{Nithin Abraham}
\affiliation[IISc]
{Department of Electrical Communication Engineering, Indian Institute of Science, Bangalore 560012, India}
\author{Kenji Watanabe}
\affiliation[NIMS1]
{Research Center for Functional Materials, National Institute for Materials Science, 1-1 Namiki, Tsukuba 305-0044, Japan}
\author{Takashi Taniguchi}
\affiliation[NIMS2]
{International Center for Materials Nanoarchitectonics, National Institute for Materials Science,  1-1 Namiki, Tsukuba 305-0044, Japan}
\author{Kausik Majumdar}
\email{kausikm@iisc.ac.in}
\affiliation[IISc]
{Department of Electrical Communication Engineering, Indian Institute of Science, Bangalore 560012, India}

\title{Anomalous Stark Shift of Excitonic Complexes in Monolayer Semiconductor}

\begin{document}
\begin{abstract}
Monolayer transition metal dichalcogenide semiconductors host strongly bound two-dimensional excitonic complexes, and form an excellent platform for probing many-body physics through manipulation of Coulomb interaction. Quantum confined Stark effect is one of the routes to dynamically tune the emission line of these excitonic complexes. In this work, using a high quality graphene/hBN/WS$_2$/hBN/Au vertical heterojunction, we demonstrate for the first time, an out-of-plane electric field driven change in the sign of the Stark shift from blue to red for four different excitonic species, namely, the neutral exciton, the charged exciton (trion), the charged biexciton, and the defect-bound exciton. Such universal non-monotonic Stark shift with electric field arises from a competition between the conventional quantum confined Stark effect driven red shift and a suppressed binding energy driven anomalous blue shift of the emission lines, with the latter dominating in the low field regime. We also find that the encapsulating environment of the monolayer TMDC plays an important role in wave function spreading, and hence in determining the magnitude of the blue Stark shift. The results for neutral and charged excitonic species are in excellent agreement with calculations from Bethe-Salpeter Equation that uses seven-band per spin tight binding Hamiltonian. The findings have important implications in probing many-body interaction in the two dimension as well as in developing layered semiconductor based tunable optoelectronic devices.
\end{abstract}
\newpage
\section{Introduction}
The ability to dynamically tune the absorption and emission properties of strongly luminescent materials is a highly desired trait for realizing tunable optoelectronic devices. Stark effect\cite{Stark14} is one such phenomenon where an external electric field is used to perturb the electronic structure of the active material and usually provides a red shift in the emission line due to a reduction in the bandgap. In Quantum Confined Stark Effect (QCSE), the electric field is applied along the confinement direction of a quantum well or other nanostructures, thereby allowing a large electric field resulting in a giant shift in the emission line well beyond the binding energy of the exciton without causing field ionization \cite{miller84,kuo05,finley04,anthony10,nakaoka06p2}. QCSE finds widespread application in ultrafast optical modulators\cite{roth07,Yamanishi92,kuo06}, self electro-optic effect devices\cite{Haas93,miller86}, tunable lasers and detectors\cite{Steinmann97,Chaisakul11}. The monolayer transition metal dichalcogenides (TMDCs) are atomically thin semiconductors which host strongly bound excitonic complexes, and serve as an excellent test bed for the investigation of QCSE\cite{klein16,wang18,chakraborty19,nikolay19,massicotte18,Chakraborty17,roch18,jiang19}. The overall energy shift ($\Delta E$) of the exciton peak due to QCSE has two components. A red shift ($\Delta E_1$) arises due to the opposite movement in energy of the constituent electron state in the conduction band and the hole state in the valence band, and is given by $\Delta E_1 = -\gamma' F - \gamma'' F^2$ where $F$ is the electric field along the out of plane direction, $\gamma'$ and $\gamma''$ are respectively the corresponding dipole moment and polarizability. The presence of reflection symmetry in monolayer TMDC makes the linear term zero, and we only observe a Stark shift that is quadratic with the applied electric field. The other component of the Stark shift, which is a blue shift ($\Delta E_2$), results from a suppression of the exciton binding energy due to the opposite motion of the wave function of the constituent electron and hole within the quantum well under the electric field. For a square well, using second order perturbation theory, one obtains $\gamma'' \propto t^4$ where $t$ is the quantum well thickness \cite{singh03}. Thus, in typical quantum wells with width extending up to several nanometers, the quadratic red shift significantly dominates over the blue shift, and thus a blue Stark shift is usually not observed. Monolayer TMDCs, owing to the extremely strong confinement due to their sub-nanometer thickness, shows a relatively weak red shift due to QCSE, and forms an interesting platform for possible observation of blue Stark shift.

In this work we demonstrate a direct observation of conspicuous blue and red Stark shift in a clean graphene/hBN/WS$_2$/hBN/Au heterojunction depending on the magnitude of the electric field applied. Accordingly, the effective shift becomes non-monotonic with the magnitude of the electric field, with the blue shift dominating at the low field regime, which can be switched to a net red shift at higher electric field. The results are in excellent agreement with theoretical results obtained by solving Bethe-Salpeter equation that uses a seven-band per spin quasi-particle Hamiltonian. Further, similar tunable shift is demonstrated for other excitonic complexes including charged exciton, charged biexciton and defect-bound exciton. Such field tunable red and blue Stark shift in different excitonic complexes has not been hitherto demonstrated and provides additional Stark tunability that has direct implication in building layered material based tunable optoelectronic devices \cite{Gu19,Scuri18,Zhou20}.

\section{Results and Discussions}
\begin{figure}[]
  \centering
  \includegraphics[scale=0.5]{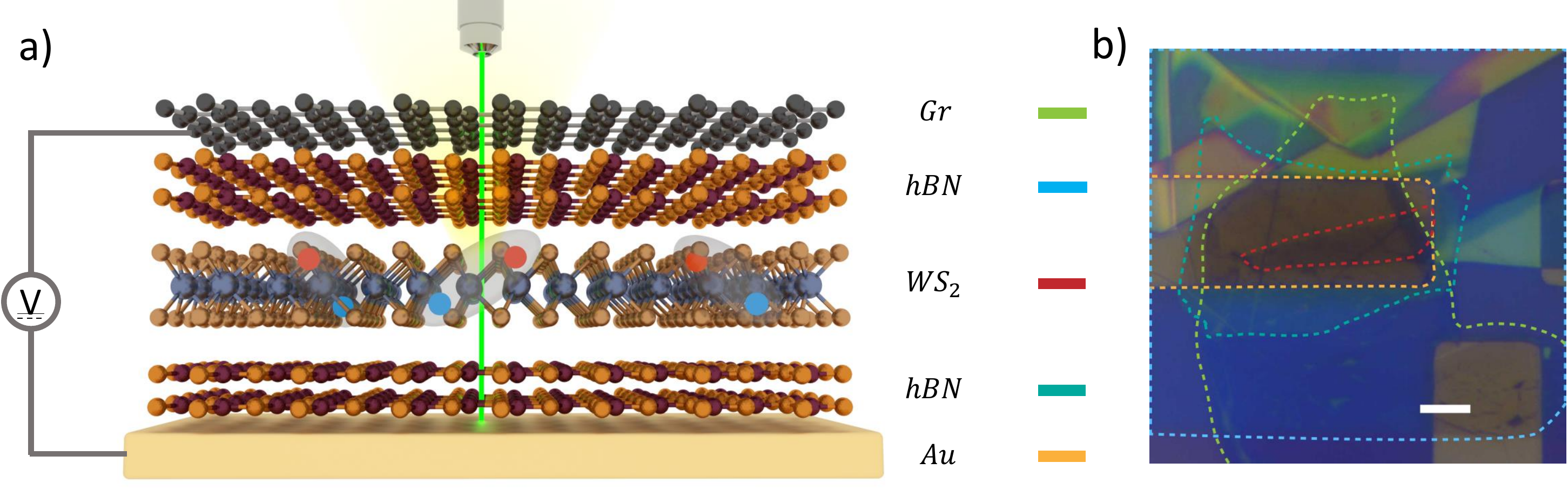}
  \caption{\label{fig:schema}\textbf{Device schematic and experimental setup:} a) Schematic representation of the graphene/hBN/WS$_2$/hBN/Au heterostructure used to investigate the effect of electric field on various multi-particle complexes. Field dependent PL spectra under a $532$ nm laser excitation is collected through the transparent top graphene electrode in a back reflection geometry. b) Optical image delineating various layers of a sample device. Scale bar is $5$ $\mu$m.}
\end{figure}

Monolayer of tungsten disulphide (1L-WS$_{2}$), which exhibits strong excitonic luminescence at room temperature, is chosen as the active material in this study. Few-layer hexagonal Boron Nitride (hBN) flakes are used to encapsulate the WS$_2$ film and form the top and the bottom dielectric layers. The hBN encapsulated stack is sandwiched between a bottom Au line and a top few-layer graphene film. The Au layer serves the dual role of the bottom electrode and the back reflector while the graphene layer acts as a transparent top electrode, as schematically illustrated in Figure \ref{fig:schema}a. The whole device fabrication has been performed by transferring layers on pre-patterned electrodes, thus avoiding any chemical processing during fabrication, which helps to retain the \txr{pristine nature of the flakes and high quality of the interfaces}. Details of the device fabrication is provided in \textbf{Methods section} and the various steps are depicted in \textbf{Supplementary Figure 1}. Five such samples D1-D5 with varying hBN thickness are fabricated and measured. All samples except D2 are annealed at $70^\circ$ C for $5$ hours at a pressure of $10^{-6}$ Torr. The optical micrograph of a typical device after completion of fabrication is shown in Figure \ref{fig:schema}b. To investigate the effect of an out-of-plane electric field, an electric field is setup between the bottom metal and the top graphene layer. The 1L-WS$_2$ layer is kept electrically floating to avoid direct carrier injection, and thus avoiding any unintentional electrostatic doping effect. The high quality of the hBN layers ensures low gate leakage current in the applied bias range, as shown in \textbf{Supplementary Figure 2} for all the samples. At each electric field, the stack is excited with a 532 nm laser and the photoluminescence spectra are collected in a back-reflection geometry. The laser power density is kept below $5$ $\mu$W/$\mu$m$^2$ ($50$ $\mu$W/$\mu$m$^2$ for sample D2) to avoid any unintentional heating of the flake.

\begin{figure}[]
  \centering
  \includegraphics[scale=0.47]{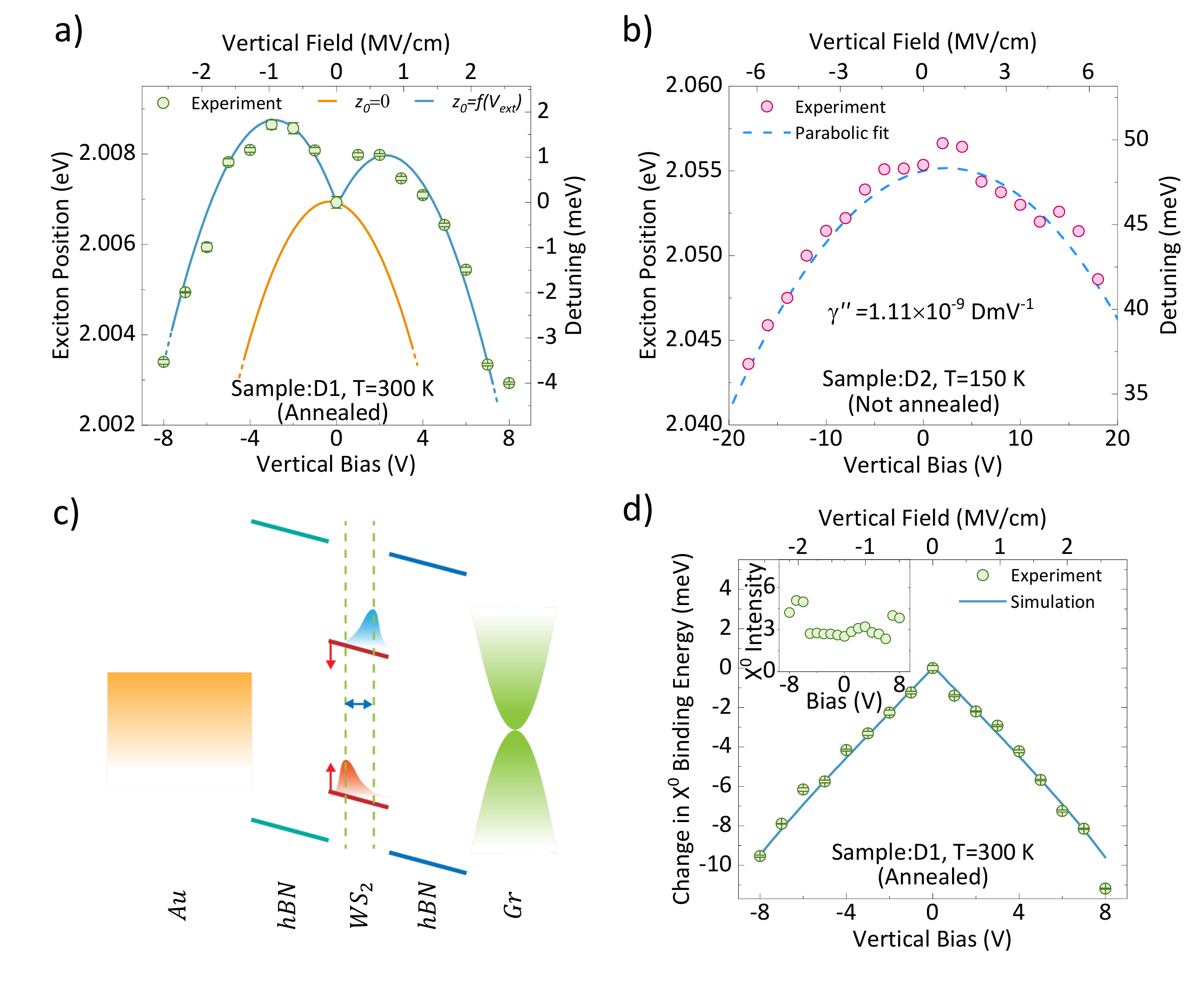}
  \caption{\label{fig:exciton}\textbf{Field dependence of exciton ($X^0$) peak:} a) Experimental data on D1 (annealed sample) at $300$ K showing blue and red shift of the peak at various field regimes along with simulation with (blue trace) and without (orange trace) considering electron-hole separation ($z_0$) induced binding energy reduction. Error bars indicate the standard deviation of the $X^0$ peak position obtained from multiple fits to the measured data. The dependence of $z_0$ on $V_{ext}$, $f(V_{ext})$ is given in \textbf{Supplementary Figure 6a.} \txr{b) Experimental data from D2 (non-annealed sample) showing absence of blue shift and a strict adherence to a parabolic red shift. A parabolic fit to the data is given by the dashed line.} c) Schematic band diagram of the device illustrating various effects determining the Stark shift of exciton. Quasi-particle levels in WS$_2$ undergo a bandgap reduction (illustrated with red arrows) leading to a red shift in the emission. Wave functions of the interacting quasi-particles experience an out of plane separation (marked with blue arrows) leading to a reduction in binding energy and hence a blue shift. d) Circles (blue trace) depicting expected binding energy reduction obtained from difference of measured (simulated) $X^0$ position and simulation with $z_0 = 0$ showing a linear reduction of binding energy with applied voltage. Error bars represent the standard deviation of the $X^0$ peak position obtained from multiple fits to the measured data. Inset: Measured PL emission intensity of the $X^0$ peak as a function of the applied voltage.}
\end{figure}

The symbols in Figure \ref{fig:exciton}a depict the variation in the position of the neutral exciton ($X^0$) from sample D1 as a function of the applied bias ($V_{ext}$) at the top electrode with respect to the bottom electrode at $300$ K. The electric field seen by the 1L-WS$_2$ is found out by assuming the continuity of out of plane component of the electric displacement vector across the hBN/WS$_2$/hBN quantum well. The right axis shows the relative shift in $X^0$ position from $V_{ext}=0$ V. \txr{To accurately capture the spectral features of the constituent peaks, the experimentally obtained spectra are fitted with Voigt profiles as shown in \textbf{Supplementary Figure 3}. The error from multiple such fits is given as the error bar in Figure \ref{fig:exciton}a and is negligible compared to the observed shifts.} The overall shift remains symmetric about $V_{ext}=0$ V. The shift in $X^0$ position initially shows a blue shift with a maximum shift of $1.7$ meV occurring at $|V_{ext}| \approx 2$ V. With $|V_{ext}| > 2$ V, the feature turns around, with a net red shift appearing at larger $|V_{ext}|$. \txr{To support the repeatability of the observations, results from another sample (D4) is given in \textbf{Supplementary Figure 4a} exhibiting a similar trend.}

\txr{Such a trend contrasts with the previous reports on QCSE in monolayer TMDCs\cite{klein16,roch18,Verzhbitskiy19} where the spectral peaks exhibit a parabolic red shift with applied field. To reproduce the observations from the existing literature, we measure the field modulation of $X^0$ peak from sample D2, which, similar to the samples from the previous reports, is not vacuum annealed at any point during the fabrication. Measured $X^0$ peak positions as a function of $V_{ext}$ are given in Figure \ref{fig:exciton}b. The data exhibits a parabolic red shift in conformance with the expected second order Stark shift. The dissimilarity between Figure \ref{fig:exciton}a and \ref{fig:exciton}b hints at the importance of the vacuum annealing step for observing the reported anomalous shift which will be discussed later.}

\txr{For sample D1 in Figure 2a, the red shift at higher $|V_{ext}|$ is expected due to QCSE which results from a reduction in bandgap arising from a change in the electronic band structure driven by the vertical electric field\cite{miller84}. However, the observations at small $|V_{ext}|$ clearly differ from conventional QCSE. The anomalous blue shift at small $|V_{ext}|$ is a purely excitonic effect and is a consequence of the redistribution of electron and hole quasiparticle wave functions constituting the exciton\cite{miller84}. The electric field creates an out-of-plane separation between the electron and the hole wave functions, generating a static dipole moment along the vertical direction as schematically illustrated in Figure \ref{fig:exciton}c. The extent of this separation is determined by the strength of the confinement. This separation reduces the attractive interaction between the electron and the hole, leading to a reduction in the binding energy of the exciton with an increasing field, which manifests as the observed blue shift.}

\txr{To obtain physical insights,} we first evaluate the electric field dependent bandgap at the $K (K^\prime)$ point in monolayer WS$_2$ using a seven-band per spin tight binding Hamiltonian \cite{Rostami13} where tungsten d orbitals and sulphur p orbitals form the basis. The details of the Hamiltonian are discussed in \textbf{Supplementary Methods S1.1}. Next, in order to include the excitonic effect, we form the two-particle exciton Hamiltonian and obtain the energy eigenvalues at the zero center of mass momentum ($\vec{Q}=\vec{0}$) by solving Bethe-Salpeter equation (BSE)\cite{Fengcheng15}. The details of the calculations are provided in  \textbf{Supplementary Methods S1.2}. An external bias dependent separation ($z_0(V_{ext})$) between electron and hole wave functions in the out of plane direction modifies the Rytova-Keldysh interaction potential\cite{Rytova67,Keldysh79} in the BSE as\cite{Tuan18,Meckbach18,Cavalcante18}
\begin{equation}\label{eq:potential}
\tilde{V}_{\vec{q},V_{ext}}=\frac{2\pi {e}^2 e^{-qz_0}}{q}\frac{1}{\epsilon(q)}
\end{equation}
where, for the case of identical dielectric environment at the top and the bottom of the monolayer,
\begin{equation}\label{eq:epsilon}
\epsilon(q)=\frac{(1+p e^{-\eta qd})\kappa}{(1-p e^{-\eta qd})}+r_0 q e^{-qz_0}.
\end{equation}
Here, $e$ is the electronic charge, $p=(\epsilon_{env}-\kappa)/(\epsilon_{env}+\kappa)$, $\eta=\sqrt{\epsilon_{\parallel}/\epsilon_{\perp}}$, $\kappa=\sqrt{\epsilon_{\parallel}\epsilon_{\perp}}$, $d$ is the thickness of the monolayer, $r_0$ is a measure of the screening length\cite{Zhang142}, $\epsilon_{env}$ is the dielectric constant of the environment, and $\epsilon_{\parallel}$ and $\epsilon_{\perp}$ are respectively the in-plane and out-of-plane dielectric constants of the TMDC. Choice of the dielectric constants\cite{Laturia18} are detailed in \textbf{Supplementary Methods S1.2}. The separation $z_0(V_{ext})$ is estimated for each applied bias as $z_0(V_{ext})=\abs{\expval{\hat{z}}{\Psi_e(V_{ext})}-\expval{\hat{z}}{\Psi_h(V_{ext})}}$ where $\Psi_{e/h}(V_{ext})$ are the electron or hole wave functions in the presence of the applied bias $V_{ext}$, as obtained from solving one-dimensional Schrodinger equation along the out-of-plane direction, and $\hat{z}$ is the position operator along the out of plane direction of the TMDC. To account for the small asymmetry about $V_{ext}=0$ in the experimental data, air gaps with thickness of few angstroms are added at the top and the bottom interfaces between TMDC and hBN layers in the simulation and is used as a fitting parameter. The orange trace in Figure \ref{fig:exciton}a is obtained by setting $z_0(V_{ext})=0$, which corresponds to the case without any electric field dependent modification on the interaction potential, and follows a quadratic red shift, as expected. For $z_0(V_{ext})\neq 0$, the interaction potential is modified according to Equation \ref{eq:potential} and hence the binding energy of the exciton reduces leading to the blue trace for the exciton eigen energies in Figure \ref{fig:exciton}a - in good agreement with the experimental observations.

\txr{In order to estimate the out of plane polarizability $\gamma''$ of $X^0$ peak with (D1) and without (D2) vacuum annealing step, we first calculate the vertical electric field $F$ seen by the TMDC layer as
\begin{equation}\label{eq:field}
F=\frac{V_{ext}}{d+d_{hBN}\frac{\epsilon_{\perp,0,TMD}}{\epsilon_{\perp,0,hBN}}}
\end{equation}
where $d_{hBN}$ is the total thickness of the top and the bottom hBN layers, $\epsilon_{\perp,0,TMD} (=6.3)$ and $\epsilon_{\perp,0,hBN}(=3.76)$ are the static out-of-plane dielectric constants of 1L-WS$_2$ and  hBN, respectively \cite{Laturia18}. $F$ is shown in the top axes of Figure \ref{fig:exciton}a, \ref{fig:exciton}b and \ref{fig:exciton}d, and is less compared to $\frac{V_{ext}}{d+d_{hBN}}$ owing to the difference between $\epsilon_{\perp,0,TMD}$ and $\epsilon_{\perp,0,hBN}$.
A parabolic fit (dashed line in Figure \ref{fig:exciton}b) models the data from D2 accurately and reveals a $\gamma''=1.11\times10^{-9}$ Dm/V\cite{klein16,roch18,Verzhbitskiy19}. At the same time, a second order fit to the experimental data at high electric field points from D1 results in an estimated effective $\gamma''$ of $3.72\times 10^{-9}$ Dm/V. The clear departure from a parabolic trend for D1 points to the deviation of the experimental data from a second order relationship expected from a pure quasiparticle effect\cite{Stark14} and hints  at the presence of a stronger electric field induced suppression of exciton binding energy for D1.}

\txr{The blue Stark shift and the larger value of $\gamma''$ in D1 as opposed to D2 are directly linked to the vacuum annealing step employed as well as the choice of materials. Transferred layered heterojunctions inevitably contain unintentional gaps at the interface between them. \txb{These gaps originate from air pockets and trapped organic polymer residues from the transfer process, both of which have low dielectric constants.} Slow vacuum annealing was proposed to mitigate this effect\cite{Tongay14} and also to achieve a cleaner interface\cite{Zhang14}. The \txb{interface} gap relaxes the net electric field seen by the monolayer TMDC due to the lower dielectric constant of \txb{trapped material} (see equation \ref{eq:field}) and manifests as a reduced polarizability. This effect can be quite significant depending on the thickness of \txc{interface} gap. Presence of the gap also introduces a larger energy barrier (compared to hBN) for the quasiparticles in the TMDC. This blocks an easy displacement of the quasiparticle wavefunctions in the out of plane direction, which is essential for the observation of the reported blue shift and to achieve a larger polarizability\cite{Maize11}. The vacuum annealing step either reduces or eliminates the gap, as is the case for D1. Apart from the annealing step, the choice of the combination of the TMDC and the barrier layer is also crucial. hBN provides lower band offsets to 1L-WS$_2$ compared with gate dielectrics like Al$_2$O$_3$ or SiO$_2$. hBN also provides a superior interface quality as well as a lower gate leakage current density which are critical for reliable observations.}

The change in the exciton binding energy with field is obtained in Figure \ref{fig:exciton}d as the deviation of experimental data to the expected parabolic trend obtained by setting $z_0(V_{ext})=0$ in the interaction potential. The blue trace in the same figure shows the difference between blue and orange traces in Figure \ref{fig:exciton}a. The linear but small variation of $z_0$ with $V_{ext}$ shown in \textbf{Supplementary Figure 6a} suggests strong confinement and that we broadly remain in the QCSE regime over the entire range of applied $V_{ext}$. This allows conventional QCSE which shows a parabolic nature to dominate at higher field regimes. This is in good agreement with the experimental observation of photoluminescence intensity being a very weak function of the applied $V_{ext}$ as shown in the inset of Figure \ref{fig:exciton}d. This also agrees well with the calculated relative change in the exciton oscillator strength\cite{Gupta19} (see \textbf{Supplementary Figure 7}), which does not vary appreciably in the applied bias range.

\begin{figure}[]
  \centering
  \includegraphics[scale=0.47]{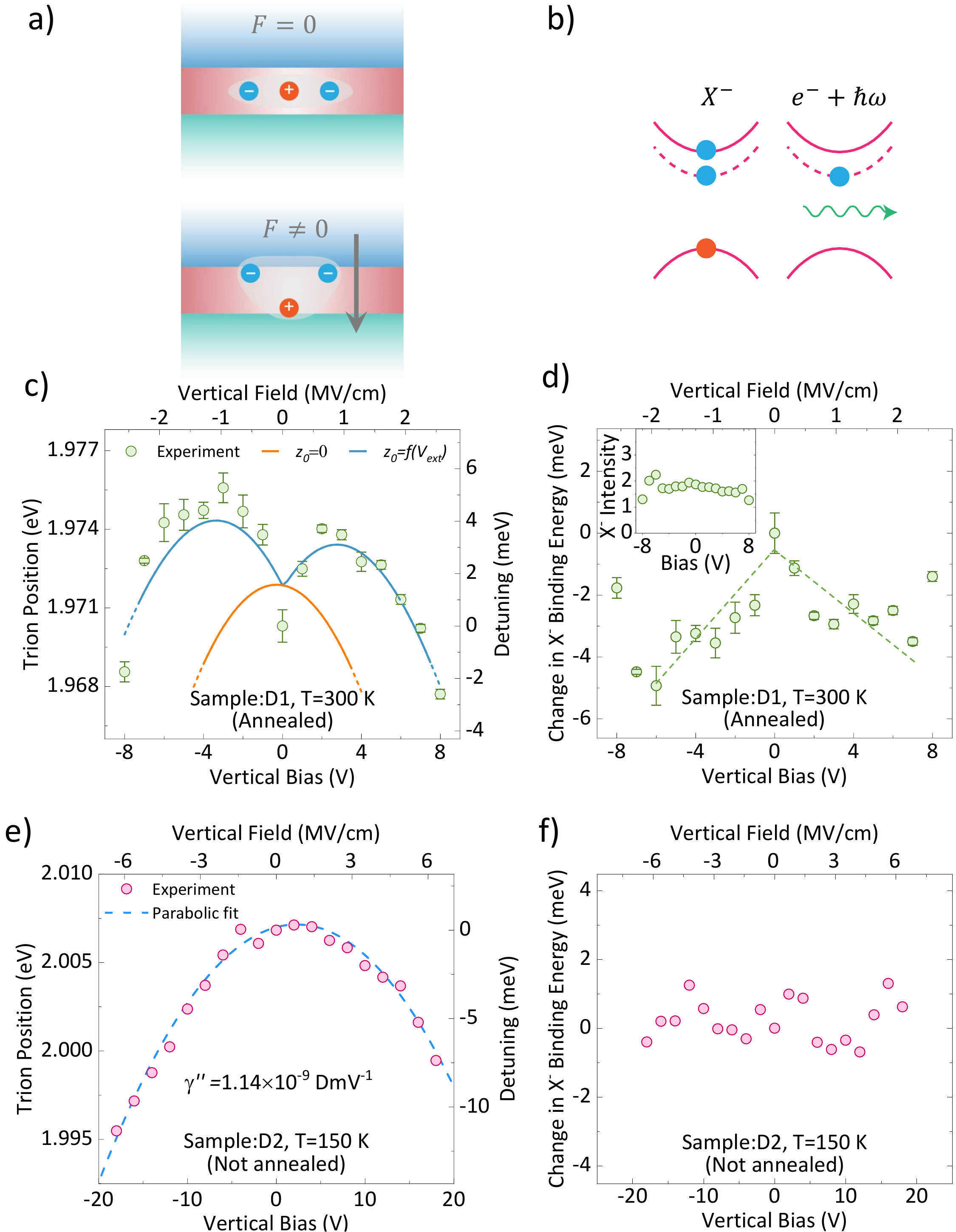}
  \caption{\label{fig:trion}\textbf{Field dependence of trion ($X^-$) peak:} a) Schematic representation of the orientation of $X^-$ in the absence (top panel) and presence (bottom panel) of a vertical field $F$. b) Valley configuration and recombination mechanism for a singlet $X^-$ state in 1L-WS$_2$. c) Experimental peak position of $X^-$ from D1 (annealed sample) at different voltages showing blue as well as red shift along with simulation with ($z_0=f(V_{ext})$, blue trace) and without ($z_0=0$, orange trace) considering binding energy reduction. Error bars indicate the standard deviation of the $X^-$ peak position obtained from multiple fits to the measured data. The dependence of $z_0$ on $V_{ext}$, $f(V_{ext})$ is given in \textbf{Supplementary Figure 6a.} \txr{d) Circles (dashed line) representing experimental data (guide to eye) on the modulation of trion binding energy obtained from the separation of $X^0$ and $X^-$ peak following equation 4. Error bars represent the standard deviation of the trion binding energy obtained from multiple fits to measured data. Inset: Dependence of $X^-$ PL emission intensity on vertical field. e) Experimental $X^-$ peak position as a function of $V_{ext}$ from D2 (non-annealed sample) exhibiting only red shift. A parabolic fit (dashed line) models the data accurately. f) Invariance of $X^-$ binding energy with changing $V_{ext}$ for non-annealed sample D2 supporting the absence of blue shift.}}
\end{figure}

To extend the study beyond two particle complexes, we further investigate the effect for higher order excitonic complexes and the resulting anomalous blue shift. The distribution of the constituent particles of a negatively charged exciton or trion ($X^-$), with and without a vertical field, is schematically illustrated in Figure \ref{fig:trion}a along with a depiction of the recombination mechanism for trion in Figure \ref{fig:trion}b. The trion qualitatively exhibits similar Stark shift as the neutral exciton, though the magnitude of the blue shift is larger, $\sim 5.2$ meV. Simulation results with (blue trace) and without (orange trace) considering the field dependent modified coulomb interaction are also shown in Figure \ref{fig:trion}c. Three particle interactions are accounted for by using a modified BSE\cite{Tempelaar19,Druppel17}, as explained in \textbf{Supplementary Methods S1.3} along with a bias dependent electron-hole interaction as given in Equation \ref{eq:potential}. The precision of the simulation for trion was limited by the prohibitively large matrix sizes needed for trion simulation. Nonetheless, the qualitative agreement with the experimental results verifies a similar origin for the blue shift in trion as the neutral exciton.

The separation ($\Delta \varepsilon$) between the $X^0$ and the $X^-$ peaks is the trion dissociation energy, and is given by \cite{Mak13,Ainane89,Huard00}
\begin{equation}\label{eq:trion}
\Delta \varepsilon (V_{ext}) = \varepsilon_{t} (V_{ext}) + \delta \varepsilon_n
\end{equation}
where $\varepsilon_{t}$ is the binding energy of the trion, and $\delta \varepsilon_n$ is the energy required for the excess electron to move to an empty state in the conduction band. $\delta \varepsilon_n$ is primarily dependent on the doping induced Pauli blocking \cite{Kallatt19}. In the current structure, where WS$_2$ remains floating, we do not electrostatically dope the system by the external voltage and thus $\delta \varepsilon_n$ plays the role of an additive constant. The change in $\Delta \varepsilon$ thus provides a direct measure of the reduction in trion binding energy at different $V_{ext}$. \txr{We plot the change in measured $\Delta \varepsilon$ and thereby the modulation of $\varepsilon_{t}$ in Figure \ref{fig:trion}d as a function of $V_{ext}$, indicating a strong suppression of the trion binding energy with increasing $|V_{ext}|$}. Note that similar to the neutral exciton, the trion intensity remains a weak function of $V_{ext}$ (see insets of Figure \ref{fig:exciton}d and Figure \ref{fig:trion}d), suggesting the absence of doping effect in the observations. \txr{Similar results from D4 are shown in \textbf{Supplementary Figure 4b} and \textbf{4c} indicating the repeatability of our observations. The role of vacuum annealing in observing the blue shift is reiterated by the purely parabolic red shift with $\gamma'' = 1.14\times10^{-9}$ Dm/V observed for $X^-$ from the non-annealed sample D2 given in Figure \ref{fig:trion}e. The stronger confinement of the electron and the hole wave functions by the \txb{interface} gap results in negligible change in the trion binding energy induced by the field, which is evident from Figure \ref{fig:trion}f.}

\begin{figure}[]
  \centering
  \includegraphics[scale=0.5]{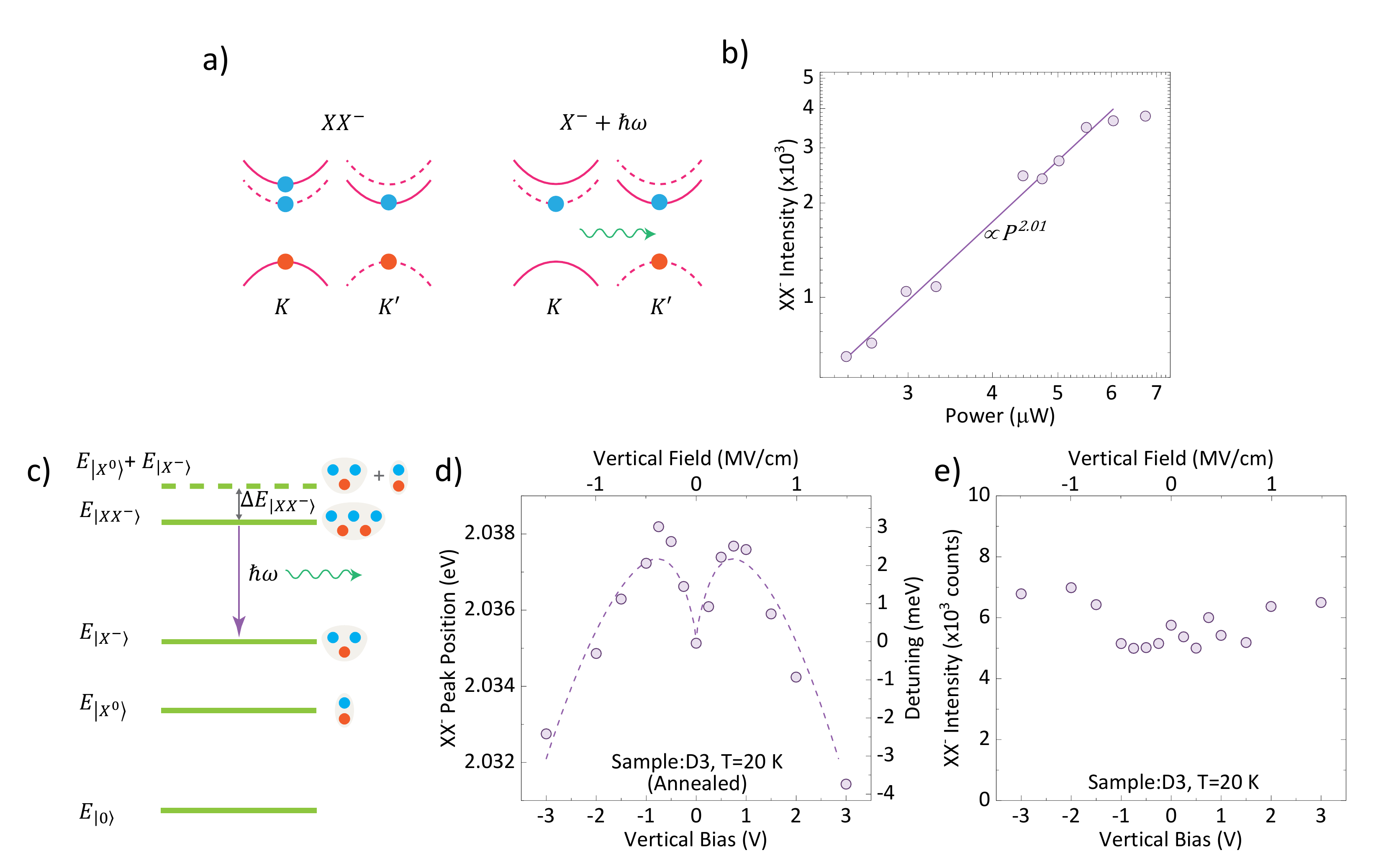}
  \caption{\label{fig:biexciton}\textbf{Field dependence of charged biexciton ($XX^-$) peak:} a) Valley configuration of the $XX^-$ state and the remaining dark $X^-$ after photon emission. b)Emission intensity as a function incident laser power for charged biexciton peak with a power law dependence of $\propto P^{2.01}$.  c) Transition diagram showing the radiative decay of $XX^-$, the states involved in its formation and the binding energy of $XX^-$. d) Field dependent blue and red shift for $XX^-$ peak at $20$ K from D3 showing a similar trend as $X^0$ and $X^-$ along with a guide to eye (dashed line). e) $XX^-$ PL emission intensity from D3 being independent of bias.}
\end{figure}

A negatively charged biexciton ($XX^-$)\cite{Barbone18,Paur19} is a five-particle excitonic complex constituting of three electrons and two holes. The spin configuration of 1L-WS$_2$ bands around $K(K^\prime)$ point forces the lower energy exciton configuration to be dark. For steady state measurement, the charged biexciton is formed through a bright exciton and a dark trion, as schematically illustrated in the left panel of Figure \ref{fig:biexciton}a. The multi-particle nature of the peak is confirmed by obtaining a quadratic ($\propto P^{2.01}$) variation of PL intensity with applied optical power (Figure \ref{fig:biexciton}b). Note that, when a $XX^-$ emits a photon, the final state is a dark trion (right panel of Figure \ref{fig:biexciton}a). Neglecting the recoil energies, the transition diagram in Figure \ref{fig:biexciton}c suggests that the energy of the photon emitted during a $XX^-$ recombination is given by
\begin{equation}\label{eq:biexciton}
\hbar \omega = E_{\ket{X^0}} - \Delta E_{\ket{XX^-}}
\end{equation}
where $E_{\ket{X^0}}$ is the neutral exciton emission energy and $\Delta E_{\ket{XX^-}}$ is the $XX^-$ binding energy. The measured Stark shift with $|V_{ext}|$ for $XX^-$ at $20$ K on sample D3 is given in Figure \ref{fig:biexciton}d with a maximum blue shift of $\sim 3$ meV. \txr{The corresponding PL spectra along with the Voigt profile fits are given in \textbf{Supplementary Figure 8}}. Equation \ref{eq:biexciton} suggests that the blue shift for the $XX^-$ peak has two origins. A blue shift in the $E_{\ket{X^0}}$ as explained earlier, and a reduction in $\Delta E_{\ket{XX^-}}$. Along with the anomalous blue shift, like the neutral and charged exciton, we also observe a strong red shift of the $XX^-$ emission peak at higher $|V_{ext}|$ due to QCSE induced quasiparticle bandgap reduction. The intensity of the $XX^-$ peak as a function of the applied $V_{ext}$ is given in Figure \ref{fig:biexciton}e. The $XX^-$ intensity being independent of the bias further confirms a constant doping in WS$_2$ throughout the bias range. The similar trend in the spectral shift of various excitonic complexes points to the universal nature for the phenomenon.

\begin{figure}[]
  \centering	
  \includegraphics[width=114.3mm]{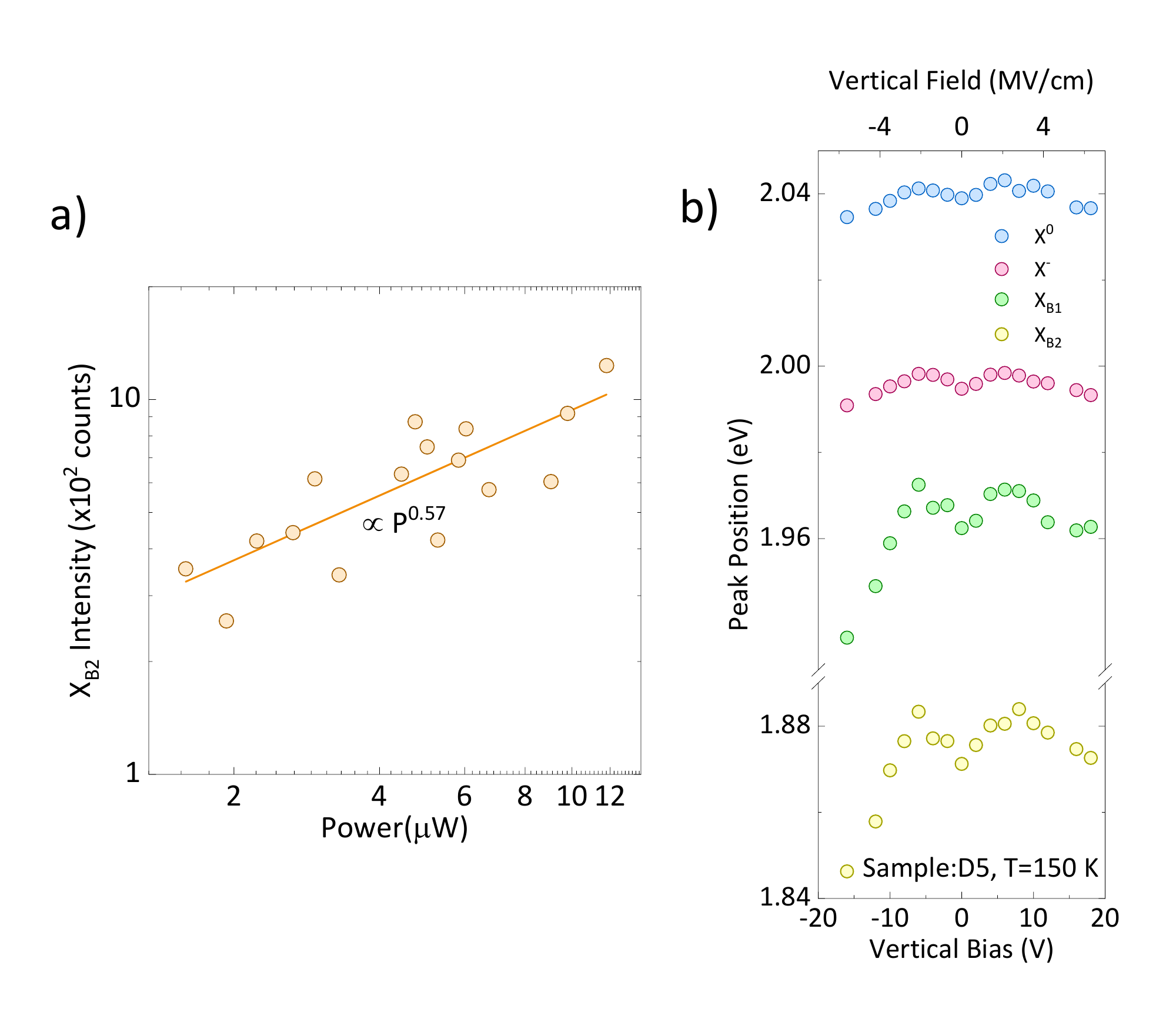}
  \caption{\label{fig:defcpos}\textbf{Verification and field dependence of defect bound exciton:} a) Variation of the PL intensity of a defect bound excitonic peak as a function of the incident optical power showing a sub-linear power dependence. b) \txc{Modulation of the peak positions for different excitonic species as a function of the applied vertical bias \txb{from sample D5} at 150 K showing blue and red shifts at different biasing regimes. Defect bound excitonic peaks $X_{B1}$ and $X_{B2}$ exhibit a larger field modulation compared to the free $X^0$ and $X^-$ peaks owing to an easily perturbed state of $X_{B1}$ and $X_{B2}$.}}
\end{figure}

We also probe the electric field induced peak position shift for a defect bound excitonic complex\cite{Chow15,Tongay13}. The defect bound nature of the peak is confirmed by a large inhomogeneous broadening and a sub-linear ($\propto P^{0.57}$) variation of PL intensity with incident optical power as shown in Figure \ref{fig:defcpos}a. \txc{Electric field modulation of the defect bound excitonic peaks $X_{B1}$ and $X_{B2}$ along with the $X^0$ and $X^-$ peaks are given in Figure \ref{fig:defcpos}b and exhibits a similar non-monotonic Stark shift for both polarities of voltage ranges. The peak positions at each voltage bias are extracted from Voigt profile fits to the experimental data as shown in \textbf{Supplementary Figure 9}. A relatively large blue shift of $\sim 12.7$ meV and an associated giant red shift of $\sim 37.6$ meV for peak $X_{B2}$ and a blue shift of $\sim 10.1$ meV and a red shift of $\sim 34.9$ meV for $X_{B1}$ could be attributed to an easily perturbed state of the defect bound exciton.} The universal trend with electric field, though of varying magnitude, in these various multi-particle complexes ascertains the prevalence of our proposed mechanism in the underlying physics governing these states.

In conventional QCSE structures, current techniques used to obtain blue shift are mainly electrostatic doping\cite{Yasukochi11,Chernikov15,He18} or relying on the asymmetry of the quantum well structure\cite{Gug98}. Change in doping alters the screening of the charges and in turn modulates the peak position. This can cause a large change in the intensity of the peaks due to transfer of oscillator strength from neutral exciton to charged exciton and vice-versa, which is undesirable in many applications as opposed to our observation (Figure \ref{fig:biexciton}e and insets of Figure \ref{fig:exciton}d and Figure \ref{fig:trion}d) of intensity being a weak function of the applied bias. Also, in a device with asymmetry, the external field either adds to or cancels the built-in field. Both these processes are usually asymmetric and depends on the sign of the applied voltage contrary to our observations.

\section{Conclusions}
\txr{In summary, we  demonstrated effective Stark tuning of the emission lines of different excitonic complexes in both red and blue directions through the application of a vertical electric field. The importance of a vacuum annealing step and the suitable choice of materials in reproducing these effects have been emphasized.} The ability to control the nature and the magnitude of the \txr{Stark} shift and the extendibility of the proposed mechanism to different many-body systems are attractive for layered material based optoelectronic applications, including tunable light emitting diode, tunable monolayer mirror, and tuning exciton-polariton through modulation of strong-coupling.

\section{Acknowledgements}
The authors thank Jithin and Prof. Mohan for providing helping with the vacuum annealing of the samples, and Veera Pandi for wire-bonding of the samples. K. M. acknowledges the support a grant from Indian Space Research Organization (ISRO), a grant from MHRD under STARS, grants under Ramanujan Fellowship and Nano Mission from the Department of Science and Technology (DST), Government of India, and support from MHRD, MeitY and DST Nano Mission through NNetRA. K.W. and T.T. acknowledge support from the Elemental Strategy Initiative conducted by the MEXT, Japan, Grant Number JPMXP0112101001, JSPS KAKENHI Grant Numbers JP20H00354 and the CREST(JPMJCR15F3), JST.

\section{Methods}
\subsection{Fabrication}
Bottom contact is defined by optical lithography using 360 nm UV source and AZ5214E resist spin-coated on an Si/SiO$_2$ substrate with 285 nm oxide formed by dry chlorinated thermal oxidation and forming gas annealing. 10 nm Ni and 15 nm Au are sputtered and lifted off by Acetone/iso-propyl alcohol rinse to form the bottom contact. The hBN flakes are transferred to a poly-di-methyl-siloxane (PDMS) sheet from Nitto tape. Flakes of suitable thickness are identified by optical contrast and dry transferred to the bottom metal contact. This process is repeated for monolayer WS$_{2}$, top hBN and graphene. Devices except D2 are vacuum annealed in $10^{-6}$ Torr pressure at $70\degree$C for 5 hours.
\subsection{Measurements}
The devices are wire bonded to closed cycle He cryostat and connected to Keithley 4200A-SCS parameter analyser for applying vertical field. A 532 nm laser is incident on the sample through a $\times50$ long working distance objective (numerical aperture = 0.5) and the emitted light is collected in a confocal manner and the spectrum is recorded with a spectrometer with 1800 lines/mm grating and CCD detector.

\section{Data availability}
Data is available on reasonable request from the corresponding author.

\bibliography{ref}
\newpage
\newpage 	


\section*{Supplementary material}

# S1 Supplementary Methods

## S1.1 Tight Binding (TB) Hamiltonian for ML WS$_2$ with field

The TB hamiltonian used to calculate the effect of field on quasiparticle band structure is given by [26]

$$\widehat{H}_{7B} = \begin{bmatrix} \widehat{H}_a & \widehat{H}_t & \widehat{H}_t \\ \widehat{H}_t^\dagger & \widehat{H}_b & 0 \\ \widehat{H}_t^\dagger & 0 & \widehat{H}_{b'} \end{bmatrix}$$

where,

$$\widehat{H}_a = \begin{bmatrix} U^a + A_1 & 0 & 0 \\ 0 & U^a + A_2 + \lambda_W \sigma & 0 \\ 0 & 0 & U^a + A_2 - \lambda_W \sigma \end{bmatrix}$$

$$\widehat{H}_b = \begin{bmatrix} U^b + B + \frac{\lambda_S}{2}\sigma & 0 \\ 0 & U^b + B - \frac{\lambda_S}{2}\sigma \end{bmatrix}$$

$$\widehat{H}_{b'} = \begin{bmatrix} U^{b'} + B + \frac{\lambda_S}{2}\sigma & 0 \\ 0 & U^{b'} + B - \frac{\lambda_S}{2}\sigma \end{bmatrix}$$

$$\widehat{H}_t = \begin{bmatrix} t_{11}f(\vec{k},\omega) & -e^{-i\omega}t_{11}f(\vec{k},-\omega) \\ t_{21}f(\vec{k},-\omega) & t_{22}f(\vec{k},0) \\ -t_{22}f(\vec{k},0) & -e^{i\omega}t_{21}f(\vec{k},\omega) \end{bmatrix}$$

and the overlap matrix is given by

$$\widehat{S} = \begin{bmatrix} s_{11}f(\vec{k},\omega) & -e^{-i\omega}s_{11}f(\vec{k},-\omega) \\ s_{21}f(\vec{k},-\omega) & s_{22}f(\vec{k},0) \\ -s_{22}f(\vec{k},0) & -e^{i\omega}s_{21}f(\vec{k},\omega) \end{bmatrix}$$

$$S_{7B} = \begin{bmatrix} 1 & \widehat{S} & \widehat{S} \\ \widehat{S}^\dagger & 1 & 0 \\ \widehat{S}^\dagger & 0 & 1 \end{bmatrix}$$

on a basis of $\{d_{z^2}, d_{x^2-y^2} + id_{xy}, d_{x^2-y^2} - id_{xy}, p_x^t + ip_y^t, p_x^t - ip_y^t, p_x^b + ip_y^b, p_x^b - ip_y^b\}$ where the $d$ orbitals are from tungsten and $p$ orbitals from sulphur. $U^{b,a,b'}$ are the potentials on the top S, middle W and bottom S layers respectively in the hBN/WS$_2$/hBN stack. $A_1$ is the on site energy of $d_{z^2}$, $A_2$ of $d_{x^2-y^2} + id_{xy}$ and $d_{x^2-y^2} - id_{xy}$, $B$ of $p_x^{t,b} + ip_y^{t,b}$ and $p_x^{t,b} - ip_y^{t,b}$. $\lambda_{W,S}$ are the SOC terms for W ans S. $\sigma$ is the spin. $t_{ij}$ are the hopping matrix elements. $f(\vec{k}, \omega)$ is the structure factor[26]. $s_{ij}$ are approximated as $t_{ij} \times 0.1eV^{-1}$. The simulation results for quasiparticle band structure for both spins along main symmetry directions are given in Supplementary Figure 5.

## S1.2 Bethe-Salpeter Equation (BSE)

BSE for monolayer WS$_2$ [27] is adapted to include the field dependence having the matrix elements at $\vec{k}$ for exciton with centre of mass momentum $\vec{Q}$ and formed from CBs $\rightarrow c$ and VBs $\rightarrow v$ as

$$\left\langle v, c, \vec{k}, \vec{Q}, V_{ext} \middle| H_X(V_{ext}) \middle| v', c', \vec{k}', \vec{Q}, V_{ext} \right\rangle = \delta_{vv'}\delta_{cc'}\delta_{\vec{k}\vec{k}'}(\epsilon_{(\vec{k}+\vec{Q})c}(V_{ext}) - \epsilon_{\vec{k}v}(V_{ext}))$$
$$- (D - X)_{vv'}^{cc'}(\vec{k}, \vec{k}', \vec{Q}, V_{ext})$$

with the direct interaction

$$D_{vv'}^{cc'}(\vec{k}, \vec{k}', \vec{Q}, V_{ext}) = \frac{1}{A}\tilde{V}_{\vec{k}-\vec{k}', V_{ext}} \left\langle c, \vec{k} + \vec{Q}, V_{ext} \middle| c, \vec{k}' + \vec{Q}, V_{ext} \right\rangle \left\langle v, \vec{k}', V_{ext} \middle| v, \vec{k}, V_{ext} \right\rangle$$

and exchange interaction $X$ vanishing at $\vec{Q} = 0$.Taking into account any separation between the electron and hole wave functions, the Fourier transform of the Rytova-Keldysh interaction

potential gets modified as [30-32]

$$\tilde{V}_{\vec{q},V_{ext}} = \frac{2\pi e^2 e^{-qz_0}}{q} \frac{1}{\epsilon(q)} \tag{1}$$

with,

$$\epsilon(q) = \frac{(1 - p_t p_b e^{-2\eta qd})\kappa}{(1 - p_t e^{-\eta qd})(1 - p_b e^{-\eta qd})} + r_0 q e^{-qz_0} \tag{2}$$

where $p_{t/b} = (\epsilon_{t/b} - \kappa)/(\epsilon_{t/b} + \kappa)$. For the case of identical environments on top and bottom Eq.2 reduces to

$$\epsilon(q) = \frac{(1 + p e^{-\eta qd})\kappa}{(1 - p e^{-\eta qd})} + r_0 q e^{-qz_0} \tag{3}$$

High frequency dielectric constants are used for the simulations as suggested in Ref[30]. A value of $\epsilon_{\infty\perp} = 6.3$ and $\epsilon_{\infty\parallel} = 13.6$ is used for $WS_2$ and $\epsilon_{\infty\perp} = 3.03$ and $\epsilon_{\infty\parallel} = 4.98$ is used for hBN [34]. The thickness of the monolayer, $d$ is set to 7 Å. The value of $r_0$ is approximated using [33] $r_0 = \frac{\epsilon^2 - \epsilon_{env}^2}{2\epsilon_s \epsilon_{env}} d$ and varied about it to fit the data. The simulated value of $z_0(V_{ext})$ as a function of $V_{ext}$ is given in Supplementary Figure 6a The simulations were carried out on an $N \times N$ $\vec{k}$ grid and the convergence of the simulation with increasing N is given in Supplementary Figure 6b.

## S1.3 Three particle BSE

Trion calculations are done by modifying the BSE to incorporate the effect of an additional charge on exciton [29,30] and adding the field dependence. Using the compact notation $\mathbf{v}' = (v', \vec{k_{v'}})$ and ignoring exchange interactions, matrix elements for a negative trion formed from

electrons with coordinates $\rightarrow \mathbf{c_1}, \mathbf{c_2}$ and and a hole $\rightarrow \mathbf{v}$ are given by

$$
\begin{aligned}
\langle \mathbf{v}, \mathbf{c_1}, \mathbf{c_2}, V_{ext} | H_{X^-}(V_{ext}) | \mathbf{v'}, \mathbf{c_1'}, \mathbf{c_2'}, V_{ext} \rangle &= (\epsilon_{\mathbf{c_1}}(V_{ext}) + \epsilon_{\mathbf{c_2}}(V_{ext}) - \epsilon_{\mathbf{v}}(V_{ext})) \delta_{\mathbf{c_1 c_1'}} \delta_{\mathbf{c_2 c_2'}} \delta_{\mathbf{vv'}} \\
&+ \frac{1}{A} V_{\vec{k_{c_1}} - \vec{k_{c_1'}}} \langle \mathbf{c_1}, V_{ext} | \mathbf{c_1'}, V_{ext} \rangle \langle \mathbf{c_2}, V_{ext} | \mathbf{c_2'}, V_{ext} \rangle \delta_{\mathbf{vv'}} \\
&- \frac{1}{A} \tilde{V}_{\vec{k_v} - \vec{k_{v'}}, V_{ext}} \langle \mathbf{v'}, V_{ext} | \mathbf{v}, V_{ext} \rangle \langle \mathbf{c_1}, V_{ext} | \mathbf{c_1'}, V_{ext} \rangle \delta_{\mathbf{c_2 c_2'}} \\
&- \frac{1}{A} \tilde{V}_{\vec{k_v} - \vec{k_{v'}}, V_{ext}} \langle \mathbf{v'}, V_{ext} | \mathbf{v}, V_{ext} \rangle \langle \mathbf{c_2}, V_{ext} | \mathbf{c_2'}, V_{ext} \rangle \delta_{\mathbf{c_1 c_1'}}
\end{aligned}
$$

Momentum conservation imposes the constraint $\vec{k_{c_1}} + \vec{k_{c_2}} - \vec{k_v} = \vec{Q}$ on the possible combinations of $\vec{k_{c_1}}, \vec{k_{c_2}}, \vec{k_v}$ where $\vec{Q}$ is the centre of mass momentum of the trion. Modification of the strength of interaction with applied field is introduced on to the electron-hole interactions. Electron-electron interactions are kept as is and $V_{\vec{q}}$ can be obtained from $\tilde{V}_{\vec{q}, V_{ext}}$ by setting $z_0(V_{ext}) = 0$. Size of matrices required to evaluate the trion hamiltonian explodes as N increases due to the increased freedom an additional particle bring in a three particle system for achieving momentum conservation. Hence the trion hamiltonian was evaluated over a coarse k grid of $17 \times 17$ due to the limited memory available. This could introduce inaccuracy in the predicted trion position but still suffice for a qualitative understanding of the observed trend.

# S2 Supplementary Figures

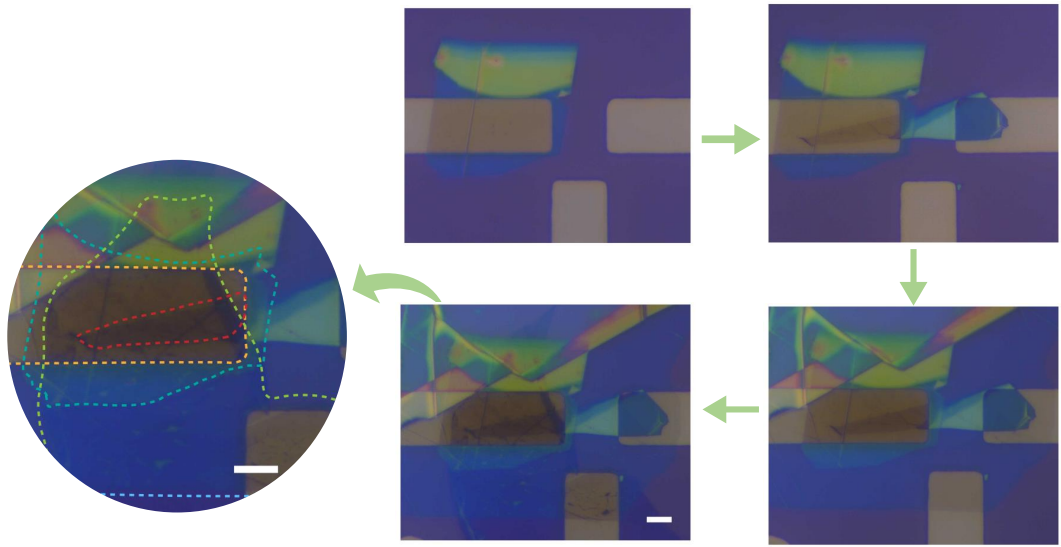

Figure 1: **Fabrication steps:** Various steps involved in the fabrication of the devices (clockwise from top) 1)Transfer of bottom hBN layer on a pre-patterned Au contact, 2) Transfer of 1L-WS$_2$, 3)Transfer of top hBN layer encapsulating the 1L-WS$_2$, 4) Transfer of thin graphene layer contacting on one pre-patterned electrode to act as top contact. 5) Zoomed in view of the junction demarcating the various layers. Scale bar is 5 $\mu$m.

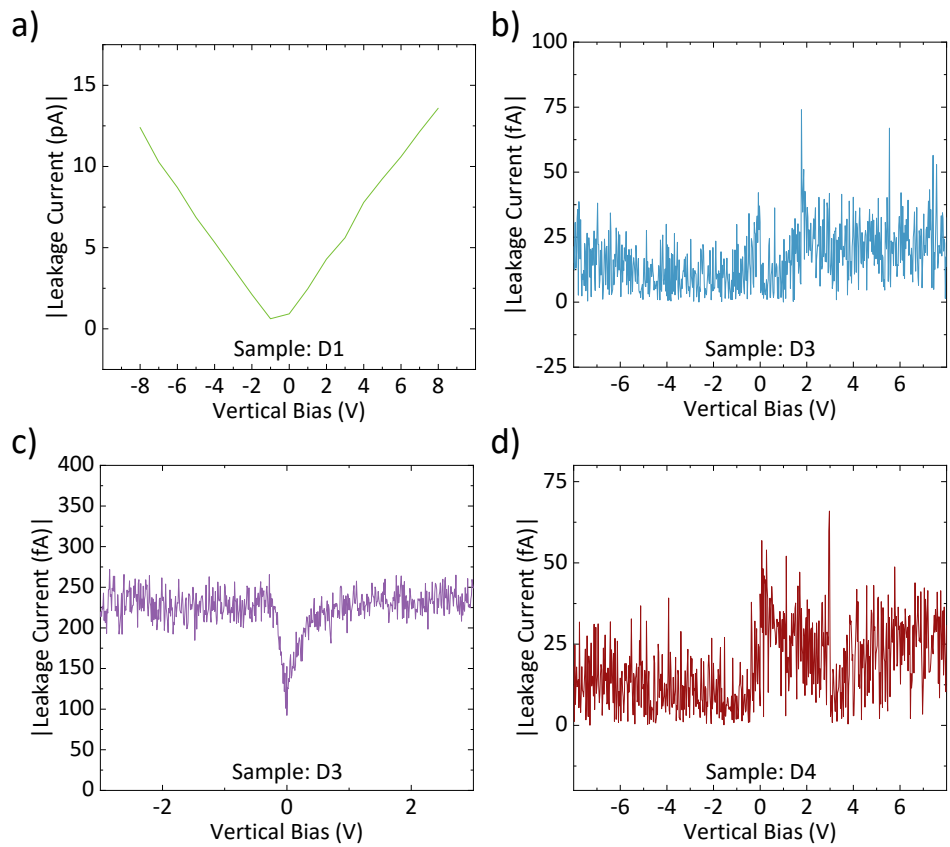

Figure 2: **Leakage current data:** Leakage current through the vertical stack as a function of the applied bias for devices D1-D4.

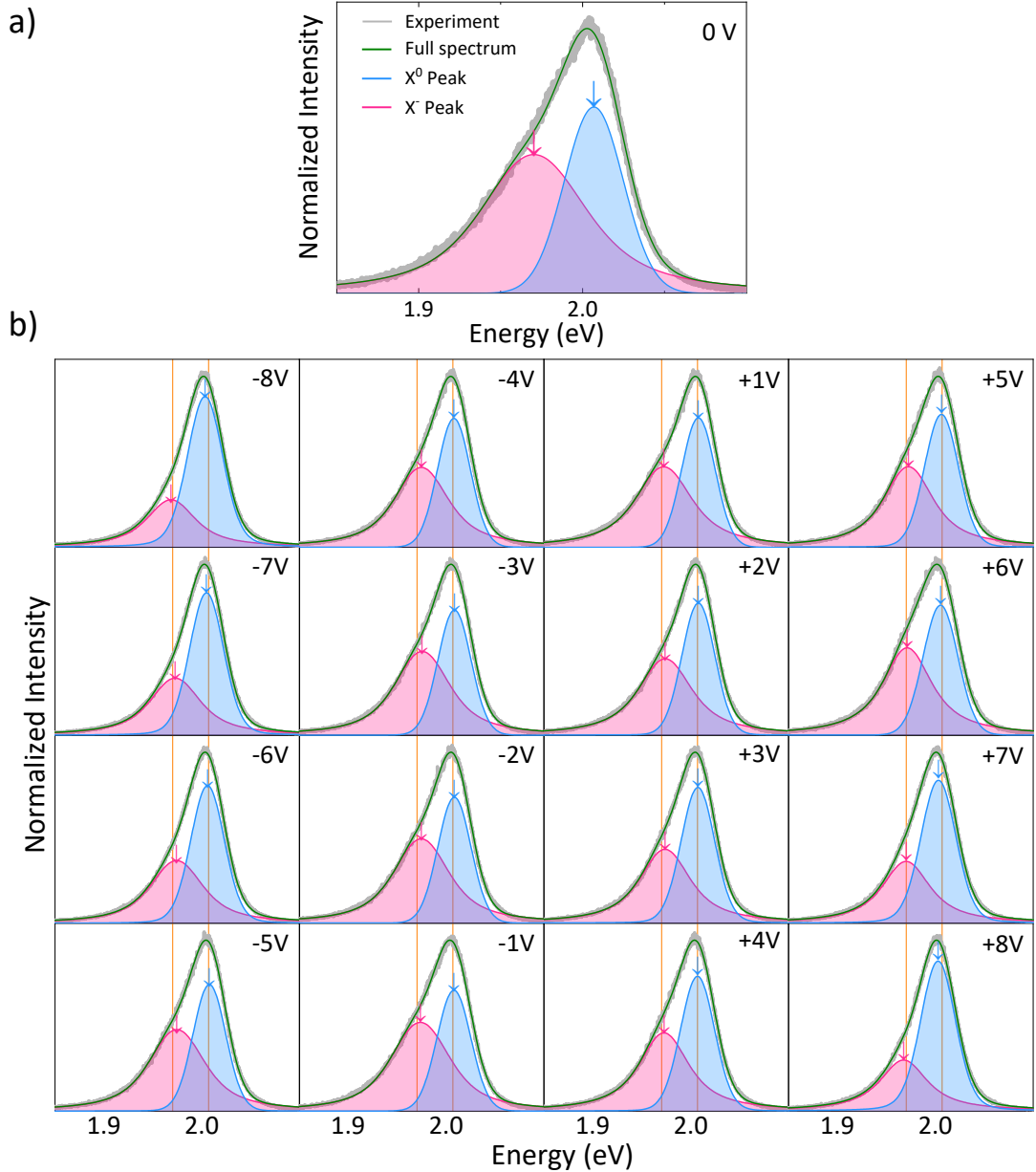

Figure 3: **Extraction of room temperature spectral features for D1:** a) To capture the features of the constituent peaks accurately, the measured (gray trace) spectra at 0 V bias from D1 at room temperature is fitted with Voigt profiles corresponding to $X^0$ (blue trace) and $X^-$ (magenta trace) emissions. The resulting green curve fits the experiment data well with a very low error. The arrows indicate the peaks of the respective curves. b) PL spectra under 532 nm illumination at different bias showing the results of Voigt profile fitting with the peak positions marked by arrows. Orange lines indicate the peak positions of $X^0$ and $X^-$ under 0 V bias.

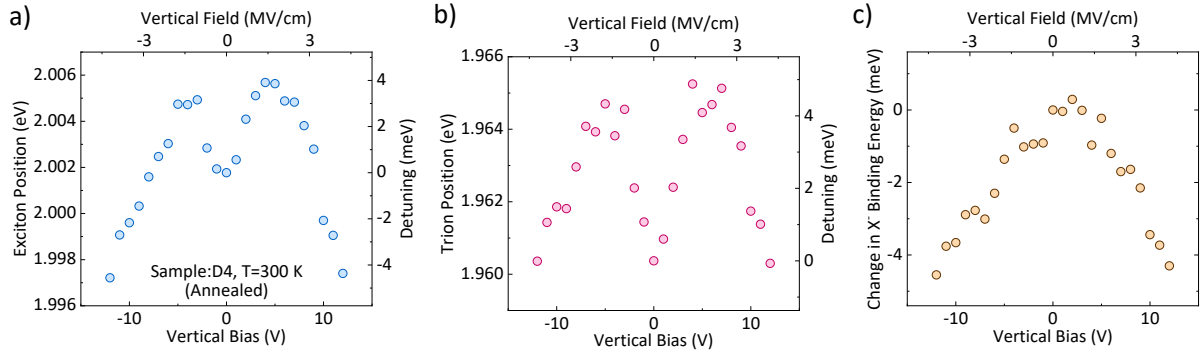

Figure 4: **Field dependent spectra from D4:** a) Exciton peak position as a function of the vertical bias with a larger blue shift obtained at room temperature. b) Trion peak position as a function of voltage bias. c) Modulation of $X^-$ binding energy with field with respect to the value at 0 V bias showing a suppression of binding energy with increasing field for the annealed device.

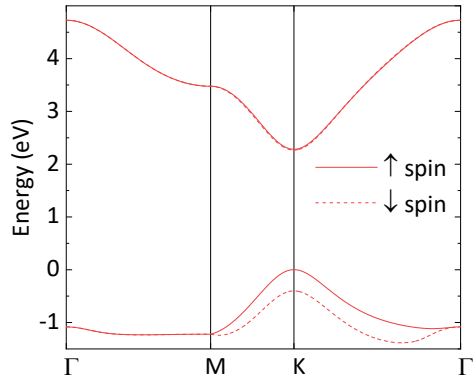

Figure 5: **WS$_2$ band structure:** Quasiparticle band structure calculated from a 7-band per spin tight binding Hamiltonian along major symmetry points with solid (dashed) lines representing spin up (down).

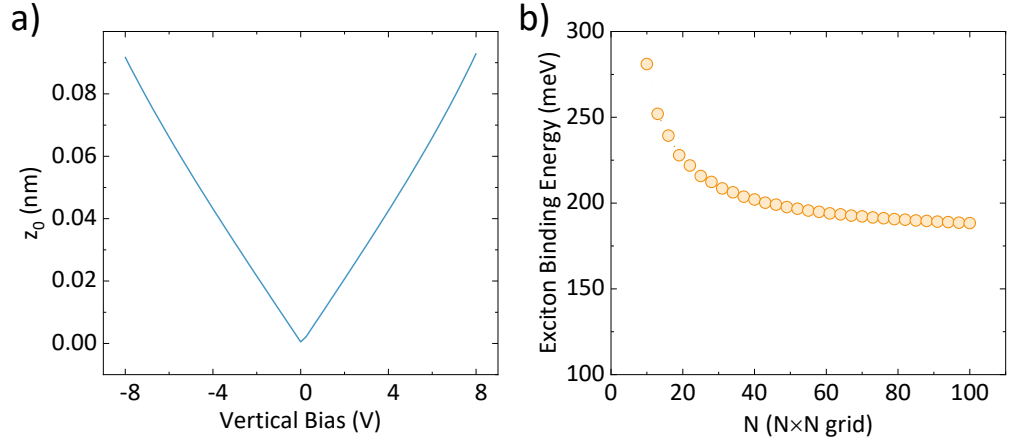

Figure 6: **e-h interaction and convergence:** a) Electron-hole separation in the out of plane direction as a function of the external bias across the quantum well. b) $X^0$ binding energy as a function of the number of mesh points showing fast convergence of the simulation to the expected value.

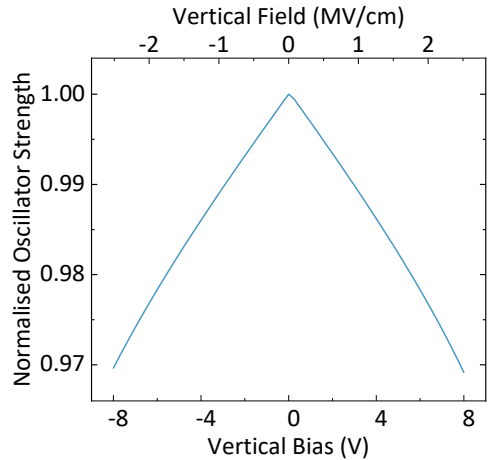

Figure 7: **Field dependent oscillator strength:** Simulated variation of the normalized oscillator strength as a function of vertical field incorporating the out of plane separation. The relatively weak field dependence of the simulated values and the small change in experimentally observed $X^0$ intensity suggest that we reamin in the QCSE regime.

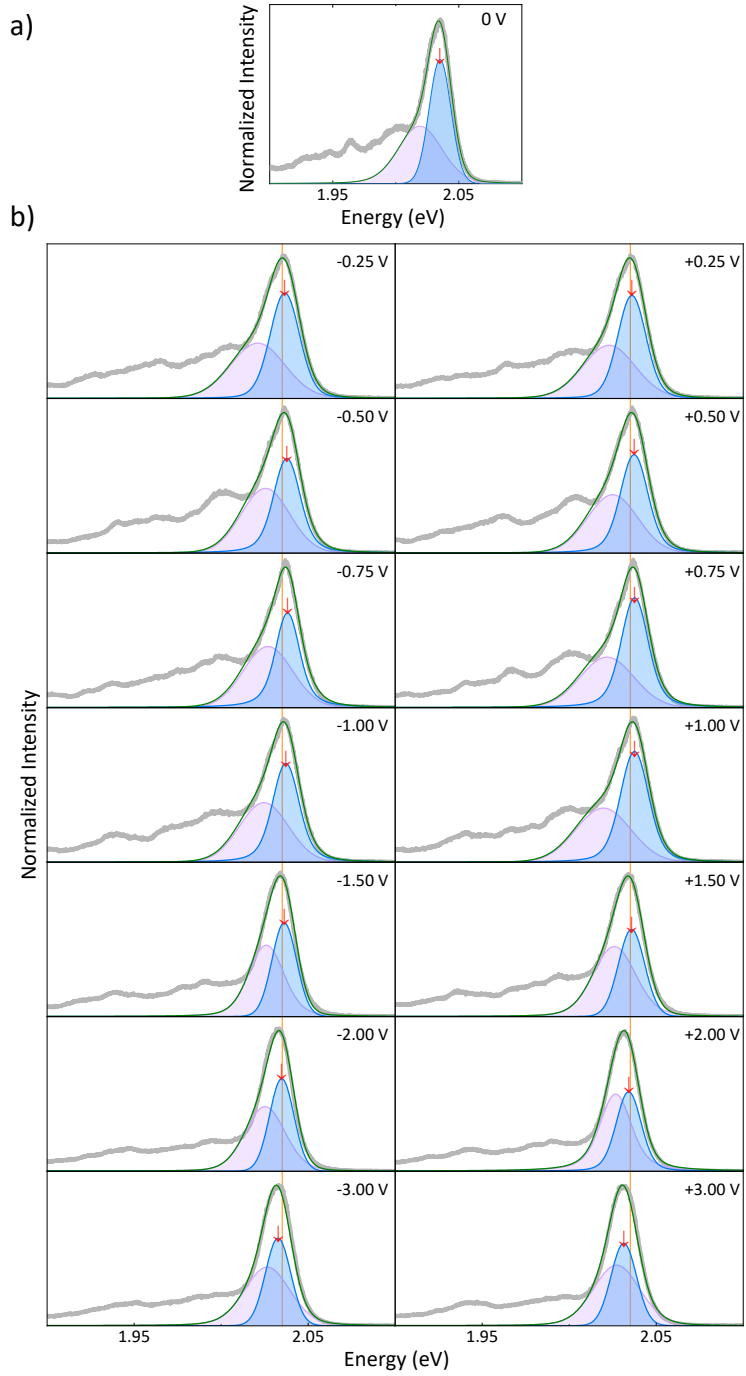

Figure 8: **Extraction of $XX^-$ spectral features:** a) Measured (gray trace) PL spectra from D4 at 20 K and 0 V bias is fitted with Voigt profiles to capture the spectral features corresponding to $XX^-$ emission (blue trace). Magenta trace accounts for the contributions from low energy defect peaks which are optically active at low temperatures. $XX^-$ peak position is marked by the arrow. b) Raw spectra and Voigt profile fitting for spectra collected at different voltage bias values along with $XX^-$ peak marked by arrows. Orange line corresponds to $XX^-$ peak position at 0 V bias.

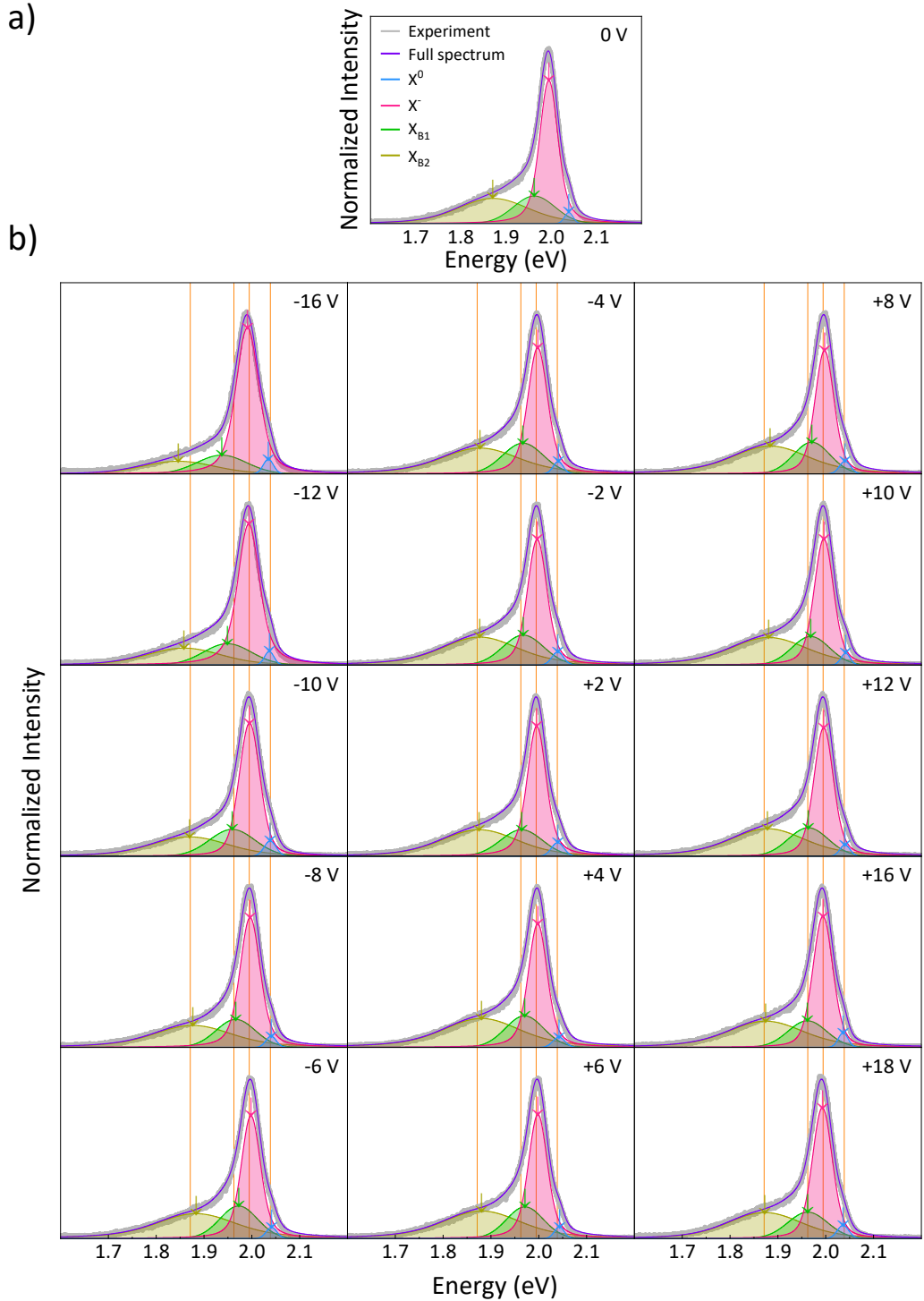

Figure 9: **Extraction of spectral features:** a) Measured spectra (gray trace) at 0 V bias at 150 K along with the Voigt profile fits corresponding to $X^0$ (blue trace), $X^-$ (magenta trace), $X_{B1}$ (green trace) and $X_{B2}$ (yellow trace). Peak positions are marked by the arrows. b) Voigt profile fits to the experimental data at different voltage bias values showing the blue and red shifts of the different peaks. Orange lines indicate the peak positions of $X^0$, $X^-$, $X_{B1}$ and $X_{B2}$ at 0 V bias.


\end{document}